\documentclass[aps,prl,reprint,groupedaddress]{revtex4-1}
\usepackage{graphicx}
\usepackage{units}
\usepackage{subfigure}

\def\PII{{2PII }}

\begin{document}

%\preprint{}


\title{Two-Pulse Ionization Injection into Quasi-Linear Laser Wakefields}% Force line breaks with \\

\author{N. Bourgeois}
\author{J. Cowley}
\author{S. M. Hooker}
\affiliation{Department of Physics, University of Oxford, Clarendon Laboratory, Parks Road, Oxford OX1 3PU, United Kingdom}

\date{\today}% It is always \today, today,

%\tableofcontents


\date{\today}

\begin{abstract}
We describe a scheme for controlling electron injection into the quasi-linear wakefield driven by a guided drive pulse via ionization of a dopant species by a collinear injection laser pulse with a short Rayleigh range. The scheme is analyzed by particle in cell simulations which show controlled injection and acceleration of electrons to an energy of \unit[370]{MeV}, a relative energy spread of 2\%, and a normalized transverse emittance of $\unit[2.0]{\mu m}$.

This is an arXiv version of the original APS paper. It should  be cited as N. Bourgeois, J. Cowley, and S. M. Hooker, Phys. Rev. Lett. 111, 155004 (2013). APS link here: http://link.aps.org/doi/10.1103/PhysRevLett.111.155004
\end{abstract}

\pacs{52.38.Kd, 41.75.Jv, 52.65.-y}
%\keywords{}

\maketitle

Laser-driven plasma accelerators can accelerate charged particles to relativistic energies with acceleration gradients at least three orders of magnitude greater than possible in a conventional, radio-frequency device \cite{Tajima:1979}. Impressive progress has been made in recent years, with several groups reporting  the generation of beams of electrons with energies in the GeV range \cite{Leemans:2006, Karsch:2007, Ibbotson:2010, Kneip:2009, Pollock:2011, Liu:2011ex, Mo:2012, Wang:2013el}.

In most experiments to date the accelerated electrons originate in the target plasma and are trapped in a highly nonlinear plasma wakefield by wave-breaking. With careful control of the driving laser and target plasma it is possible to generate beams with relative energy spread and shot-to-shot reproducibility at the level of several percent \cite{Mangles:2004, Geddes:2004, Faure:2004, Osterhoff:2008}, but it is recognized that better control of the injection process would  significantly improve the values of, and reduce the fluctuations in, the electron bunch parameters.

Several techniques for controlling electron injection have been described, including colliding-pulse injection, \cite{Faure:2006vy} density ramp injection \cite{Bulanov:1998vc, Gonsalves:2011ds, Schmid:2010}, ionization injection \cite{Mcguffey:2010, Pak:2010} and injection controlled with an external magnetic field \cite{Vieira:2011ko}. However, to date these techniques have only been demonstrated to control injection into nonlinear wake fields.

It would be advantageous also to be able to control the injection of particles into linear or quasi-linear wakefields, which arise for driving laser pulses with a peak normalized vector potential $a_0 \lesssim 1$. The linear regime offers several advantages: the range of phases available for acceleration is equal for  positive and negative particles, allowing, for example, acceleration of electrons and positrons; self-injection is avoided, preventing unwanted dark current; relativistic self-focusing of the driving laser pulse can be avoided, allowing intensity-independent guiding of the driving laser pulse in a plasma channel and control of the particle energy gain through adjustment of the laser intensity. 

In this letter we consider an extension of the ionization injection scheme with a single pulse to the case of linear or quasi-linear wakefields. With single-pulse ionization injection, partially-ionized species within the plasma are ionised by the driving laser pulse, and these additional electrons can become trapped in the plasma wakefield. Chen et al. \cite{Chen:2012du} have recently shown that trapping does not occur for resonant Gaussian laser pulses with a peak normalized vector potential $a_0 \lesssim 1.7$; in its simplest form, therefore, ionization injection will not occur in linear wakefields. The reason for this is that ionization will principally occur on the leading edge of the driving laser pulse, but for the ionised electrons to be trapped they must be born in a phase of the plasma wave for which the plasma potential $\Phi$ is sufficiently positive. For  intense laser pulses, and hence  nonlinear wakefields, the regions of ionization and the trapping partially overlap, but for linear wakefields this is not the case.

Here we consider the use of a second, co-propagating laser pulse to inject electrons into the linear wakefield driven by the primary laser pulse; for convenience we call this approach  ``two-pulse ionization injection'' (\PII). An important feature of the \PII scheme is that the parameters of the injecting laser pulse are adjusted so that it diffracts faster than the driving laser pulse, meaning that its intensity will remain high --- and hence electron trapping will occur --- in only a localized region. In contrast, the driving laser pulse can be guided in a plasma channel or hollow capillary waveguide, allowing the trapped electrons to be accelerated to high energy \cite{Spence:2000uv, Cros:2000ux, Geddes:2005wh}.

The use of two laser pulses confers considerable advantages over ionization injection with a single pulse. Single-pulse ionization injection injects electrons over a significant fraction of the accelerator length; this leads to a large energy spread --- unless the dopant species is localised \cite{Pollock:2011, Liu:2011ex}, which introduces additional technical challenges. As shown below, for optimum delays between the driving and injection laser pulses,  the wakefields driven by the two pulses add coherently; this assists trapping since electrons are born into a large-amplitude wakefield which adiabatically evolves to a linear wakefield as the injection pulse diffracts. Further, the focal spot size of the injection laser pulse can be much smaller than that of the driving laser pulse, which ensures that electrons are injected close to the axis of the wakefield, and hence that the emittance of the generated beam is low. The charge of the injected bunch may be controlled by adjusting the spot size or the intensity of the injection pulse or the density of the dopant species. Finally, the period of the plasma wave into which electrons can be trapped may be controlled by adjusting the delay between the drive and injection pulses; trapping electrons into plasma periods further behind the driving pulse increases the energy to which electrons can be accelerated in longitudinally-tapered plasmas \cite{Rittershofer:2010}.

The scheme described here is related to the ``Trojan horse''  scheme for controlling injection into \emph{beam}-driven wakefields described by Hidding et al. \cite{Hidding:2012}. We note that the electric self-fields of a laser-driver are much greater than the self-fields of a beam-driver, and hence the intensity of the injection laser pulse in the \PII scheme must be much higher than in the Trojan horse case and, importantly, higher than that of the laser driver. As such, in the \PII scheme the injection pulse will drive --- albeit for only a short distance --- a significant wakefield of its own, changing the dynamics of injection significantly from those occurring in the Trojan horse case. 

The method described here is also related to the LILAC scheme described by Umstadter et al. \cite{Umstadter:1996, Dodd:1997} in which an injection laser pulse is used to promote injection of electrons into the wakefield produced by a driving laser pulse. Both transverse and collinear geometries have been considered. The LILAC scheme differs from the \PII scheme in that for LILAC: (i)  the mechanisms responsible for injection are considered to be modification of the motion of electrons within the wakefield, and of the wakefield itself \cite{Hemker:1998te}, by the ponderomotive force of the injection laser; (ii) the intensity of the driving pulse is relatively high ($a_0 \geq 1.5$) and hence produces a nonlinear wakefield. We note that in their work Umstadter et al. mention that photoionization could be used to control injection, but they do not analyze this case.

\begin{figure}
\includegraphics[scale=0.9]{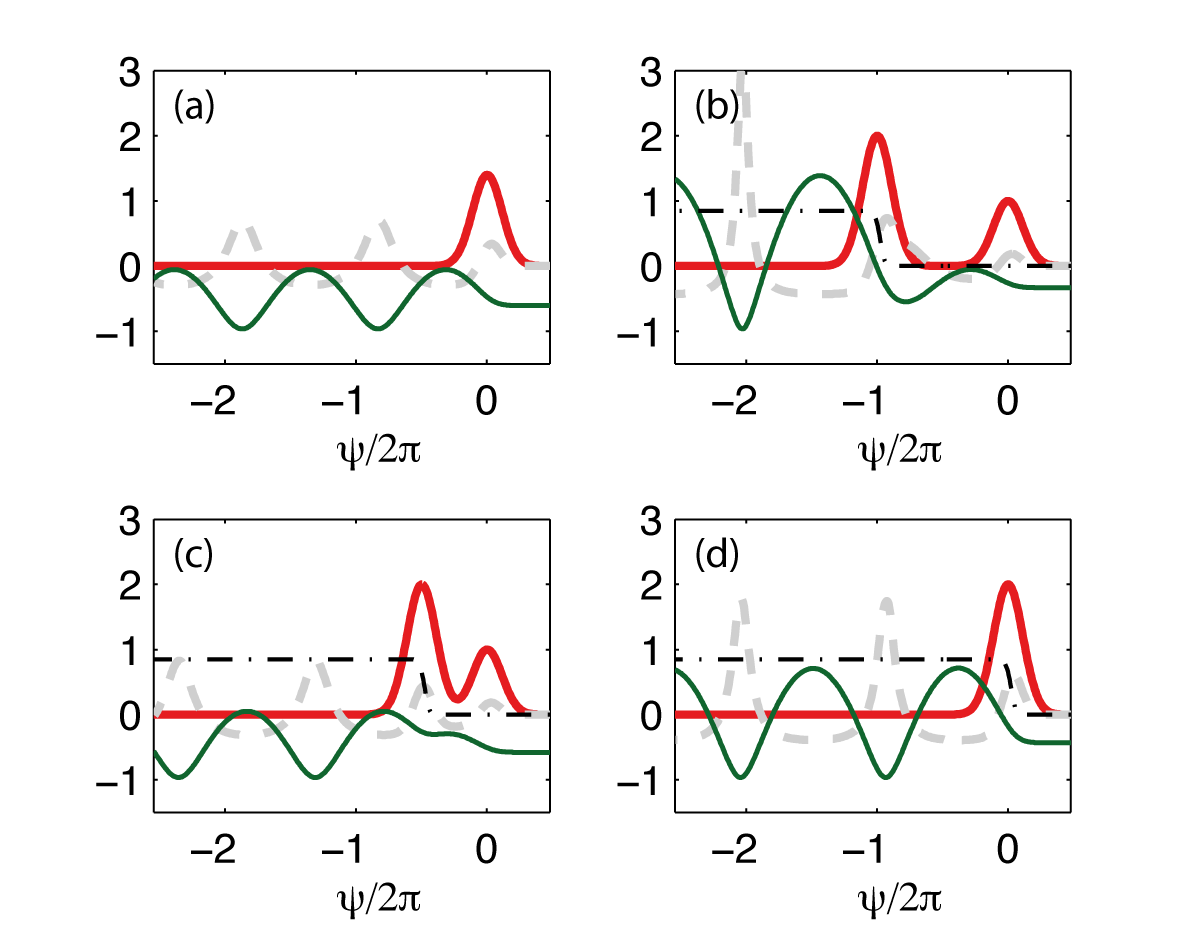}
\caption{One-dimensional fluid simulations of the electron density (gray,  dashed), trapping parameter $\Delta \mathcal{H}$ (green, thin solid), and fractional ionization of N$^{5+}$ ions (black,  dash-dot) produced by: (a) a single laser pulse with a peak normalized vector potential $a_\mathrm{drive} = 1.3$; and a pair of laser pulses with $a_0^\mathrm{drive} = 1.0$ and $a_0^\mathrm{inj} = 2$ and a relative delay of $\Delta t = 2 \pi /\omega_\mathrm{p}$ (b), $\Delta t = \pi /\omega_\mathrm{p}$ (c), and $\Delta t = 0$ (d). For all plots the laser pulses  (red, thick solid) have a Gaussian temporal profile of rms duration $\tau_\mathrm{rms} = 1/k_\mathrm{p} c$.}
\label{Fig:Fluid_sims}
\end{figure}

In order to understand the operation of the \PII  scheme  in more detail we first consider the one-dimensional (1D) motion of electrons within a wakefield. This motion is determined by the Hamiltonian $\mathcal{H}(u,\psi) = (\gamma_\perp^2 + u^2)^{1/2} - \beta_\mathrm{p} u - \phi(\psi)$, where $\gamma = (\gamma_\perp^2 + u^2)^{1/2}$ is the Lorentz factor of the electron, $u = \gamma \beta$ is the normalized longitudinal momentum, $\beta_\mathrm{p}$ is the normalized phase velocity of the plasma wave, and $\psi = k_\mathrm{p}(x - \beta_\mathrm{p}t)$. Here $k_\mathrm{p} = \omega_\mathrm{p}/c$, $\phi(\psi) = e \Phi / m_\mathrm{e} c^2$, and the angular plasma frequency is $\omega_\mathrm{p} = (n_\mathrm{e} e^2 / m_\mathrm{e} \epsilon_0)^{1/2}$.  An electron born at rest has  $\mathcal{H} = \mathcal{H}_\mathrm{i}  = 1 - \phi(\psi_\mathrm{i})$; trapping requires \cite{Chen:2012du} $\Delta \mathcal{H} = \mathcal{H}_\mathrm{s} - \mathcal{H}_\mathrm{i} > 0$ where the separatrix $\mathcal{H}_\mathrm{s} =  \gamma_\perp(\psi_\mathrm{min})/\gamma_\mathrm{p} - \phi(\psi_\mathrm{min})$ in which $\phi(\psi_\mathrm{min})$ is the potential of the plasma wave at the trapping phase  $\psi_\mathrm{min}$.

Figure \ref{Fig:Fluid_sims} shows  wakefields calculated from the 1D fluid equations, within the quasi-static approximation,  for the case of laser pulses with a Gaussian temporal profile of root-mean-square (rms) duration $\tau_\mathrm{rms} = 1/k_\mathrm{p} c$. Also shown is the  fractional ionisation of nitrogen atoms, calculated from the ADK ionization rates \cite{ADK:1986}. We choose nitrogen as a dopant since for the laser parameters considered low ionization stages will be ionized by the leading edge of the driving pulse, but N$^{n+}$ ions with $n \geq 5$ can only be ionized near the peak of the injection pulse. Figure \ref{Fig:Fluid_sims}(a) shows the wakefield driven by a \emph{single} laser pulse with a peak normalized vector potential $a_0 = 1.3$. In agreement with earlier work by Chen et al.\  \cite{Chen:2012du}, it is clear that ionized electrons cannot be trapped since $\Delta \mathcal{H} < 0$ for all phases $\psi$.

Figure\ref{Fig:Fluid_sims}(b) - (d) shows how a second, collinear laser pulse may be used to trap electrons within the wakefield of a lower intensity drive pulse. In these simulations the amplitude of the injection pulse is $a_0^\mathrm{inj} = 2$, which is sufficiently high to ionize $N^{5+}$ ions. In Figure\ref{Fig:Fluid_sims}(b), the delay between the two pulses corresponds to a phase delay  $\Delta \psi_\mathrm{inj} = 2 \pi$, and hence the wakefields driven by the two laser pulses reinforce each other. Now electrons are ionized from the dopant in regions for which $\Delta \mathcal{H} > 0$, and hence they will be trapped. Figure\ref{Fig:Fluid_sims}(c) shows that electrons will also be ionized and trapped for $\Delta \psi_\mathrm{inj} = \pi$; however, in this case the trapping is  weak since the wakefields driven by the two pulses partially cancel each other. Figure\ref{Fig:Fluid_sims}(d) shows the case for $\Delta \psi_\mathrm{inj} = 0$ and drive and injection pulses with $ a_0^\mathrm{inj} = a_0^\mathrm{drive}  = 1$, which is equivalent to a single pulse with $a_0^\mathrm{drive} = 2$. In agreement with  \cite{Chen:2012du}, this shows that the injection pulse of Figs \ref{Fig:Fluid_sims}(b) and (c) would be sufficient for ionization trapping on its own --- albeit without the advantages of greater control and localized injection of the \PII scheme --- since there exists a small window within which ionization occurs and $\Delta \mathcal{H}  > 0$.

To obtain a more detailed quantitative understanding we have performed 2D particle-in-cell simulations  using the OSIRIS code \cite{Fonseca:2002ln}. In these simulations the drive and injection laser pulses had a Gaussian temporal profile with $k_\mathrm{p} c \tau_\mathrm{rms} = 1$, and Gaussian transverse profiles of spot size ($1/e^2$ radius) $w_{0}^\mathrm{drive} = \unit[30]{\mu m}$ and $w_{0}^\mathrm{inj} = \unit[8]{\mu m}$ respectively. The two pulses  were focused at the entrance of a plasma  channel with an axial electron plasma density of $n_\mathrm{e}(0) = \unit[2.0 \times 10^{18}]{cm^{-3}}$, with a parabolic transverse density profile matched to the spot size of the driving pulse.  The boundary between the entrance to the plasma channel and vacuum took the form of a linear ramp of length $15 c/\omega_p \approx \unit[60]{\mu m}$ which was sufficiently long to avoid unwanted self injection at the vacuum-plasma boundary. Neutral nitrogen atoms, with a density equal to $5 \%$ of the initial axial electron density were distributed uniformly throughout the plasma; the nitrogen concentration was chosen to be sufficient to provide a reasonable number of trapped electrons without leading to distortion of the plasma channel following ionization of low-charge states by the driving laser pulse. A moving window of $95 \times \unit[450]{\mu m}$ was used with a $2500 \times 750$ grid, with 4 particles per cell for the plasma and 2500 particles per cell for the nitrogen atoms. The charge of the electrons ionized from the nitrogen atoms was deposited using a quadratic interpolation scheme. 

\begin{figure}[tb]
\includegraphics[scale=1.0]{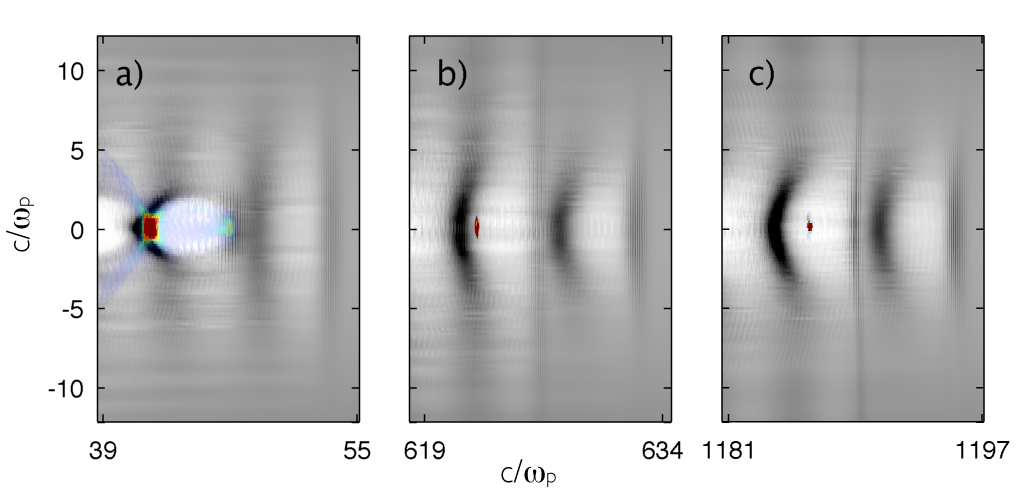}
\caption{Calculated density of electrons for (a) $z = \unit[0.2]{mm}$ and \unit[2.3]{mm} and (c) \unit[4.5]{mm}. Shown in grayscale are electrons ionized from hydrogen and N$^{< 5+}$ ions; the color scale shows electrons ionized from N$^{\geq 5+}$ ions. Laser-plasma parameters are $n_e(0) = \unit[2 \times 10^{18}]{cm^{-3}}$, injection pulse $a_0 = 2$ and drive pulse $a_0 = 1$; both laser pulses have a Gaussian temporal profile with $L = k_p c \tau_\mathrm{rms}$.}
\label{Fig:Snapshots}
\end{figure}

Figures \ref{Fig:Snapshots}(a), \ref{Fig:Snapshots}(b) and \ref{Fig:Snapshots}(c) show, at various points $z$ along the plasma, the densities of electrons in the hydrogen plasma channel and from ionization of N$^{n+}$ ions with $n\geq 5$. Figure \ref{Fig:Snapshots}(a) shows  electrons being ionized from the N$^{n+}$  ions near the axis and slipping backwards relative to the wakefield. Some of these electrons are trapped at the rear of the second plasma period behind the driving pulse --- but others continue to slip back to form a cone of ejected electrons. In Fig.\ \ref{Fig:Snapshots}(b), the intensity of the injection pulse has been reduced by diffraction sufficiently for ionization of N$^{\geq5+}$ ions to cease; a well defined electron bunch has started to form at the rear of the plasma period, with a few electrons still slipping out of this bucket in a cone of larger angle, reflecting their larger forward momentum. The electron bunch is then accelerated in the wakefield, with little further loss of electrons, as shown in Fig.\ \ref{Fig:Snapshots}(c). The total charge in the bunch is $\unit[5]{p C}$. A detailed analysis of the simulations shows that the only electrons which are trapped and accelerated are those ionized by the injection pulse from N$^{n+}$ ions with $n\geq 5$.

\begin{figure}[tb]
\includegraphics[scale=0.9]{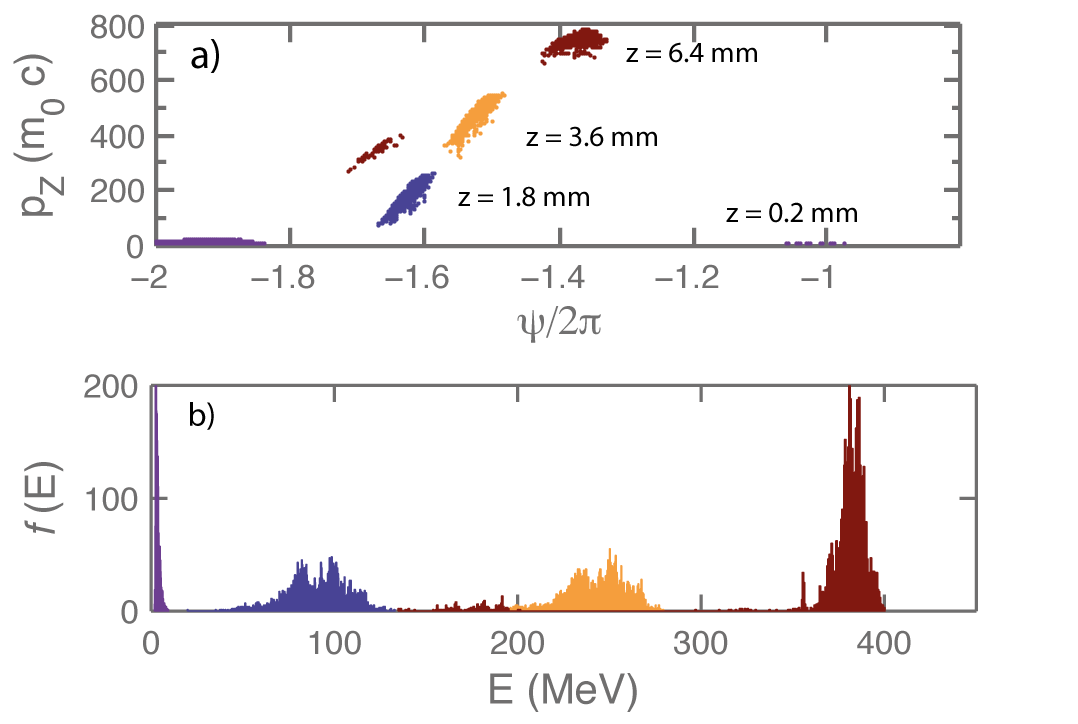}
\caption{Calculated phase space distribution (a) and energy spectrum (b) of  electrons ionized from  N$^{n+}$ ions, with $n\geq 5$,  at distances of $z = \unit[0.2]{mm}$, \unit[1.8]{mm}, \unit[3.6]{mm} and \unit[6.4]{mm} of acceleration. Only electrons remaining in the simulation box are included.}
\label{Fig:Phase_energy}
\end{figure}

Figure  \ref {Fig:Phase_energy} shows, for the same laser-plasma parameters as in Fig.\ \ref{Fig:Snapshots},  the phase space distribution and energy spectrum of electrons ionized from  N$^{n+}$ ions, with $n\geq 5$. As expected, the injected electrons are first trapped at the back of the second plasma period behind the driving pulse; they move forward relative to the wakefield as they gain energy; and phase-rotation near the point of dephasing reduces the energy spread of the trapped bunch. The maximum electron bunch energy is approximately \unit[370]{MeV}. In these simulations a second group of low energy, low charge electrons is injected close to the point of dephasing of the main bunch due to ionization by the the diffracted, but temporally compressed injection pulse; other simulations show that reducing the plasma density slightly prevents this second injection event.

\begin{figure}[tb]

\includegraphics[scale=1.0]{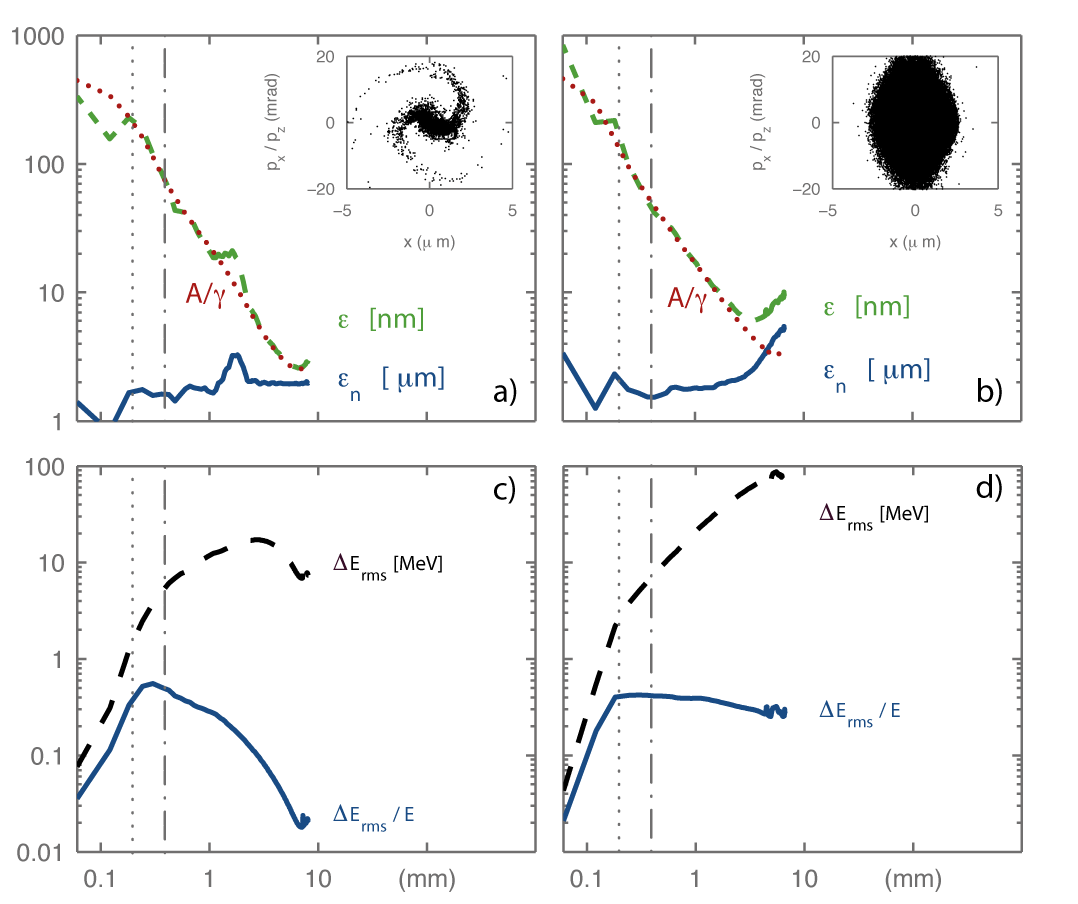}
\caption{Evolution  with position $z$ of the drive laser pulse along the plasma channel of the  parameters of electrons ionized from N$^{\geq 5+}$ for  drive and injection laser pulses with identical duration  equal to  (a, c)  $k_\mathrm{p} c \tau_\mathrm{rms} = 1$ and (b, d) $k_\mathrm{p} c \tau_\mathrm{rms} = 2.2$. Shown are the relative energy spread $\Delta E_\mathrm{rms}/E$ (blue, solid) and  $\Delta E_\mathrm{rms}$ (black, dashed), where $\Delta E_\mathrm{rms}$ is the rms energy spread, the geometric emittance $\epsilon_\mathrm{rms}$ (green, dashed), and normalised emittance $\epsilon_\mathrm{n, rms}$ (blue, solid). Also shown is a plot of $A/\gamma$ (red, dotted), where $\gamma$ is the mean relativistic factor of the trapped electrons, with the parameter $A$ adjusted to fit the minimum value of $\epsilon_\mathrm{rms}$. The inset shows the beam transverse distribution of the electron bunch at the point where the energy spread is minimum. The dotted vertical line shows where injection stops and the dotted dashed  vertical line shows where the mean velocity of the injected electrons first exceeds the phase velocity of the plasma wakefield. Only electrons in the first injected bunch which remain in the simulation box are included in calculations of these parameters.}
\label{Fig:Emittance}
\end{figure}

Figure \ref {Fig:Emittance} shows the evolution along the plasma accelerator of the emittance and energy spread of electrons ionized from N$^{\geq 5+}$  for drive ($a_0 = 1$) and injection ($a_0 = 2$) laser pulses of identical duration equal to (a, c) $k_\mathrm{p} c \tau_\mathrm{rms} = 1$ and  (b, d) $k_\mathrm{p} c \tau_\mathrm{rms} = 2.2$. The normalized rms  transverse emittance of the accelerated electron bunch is calculated \cite{Floettmann:2003} from $\epsilon_\mathrm{n, rms} = \sqrt{\left<x^2\right>\left<p_x^2\right> - \left<x p_x \right>^2}$ where $x$ is the transverse position and $p_x$ is the transverse momentum normalised to $m_e c$. Here the symbol $\left< \right>$ denotes the second central moment, i.e. $\left<x y\right> = \overline{xy} - \overline{x}~\overline{y}$ in which  the bar indicates the average over particles. The geometric emittance \cite{Floettmann:2003} is calculated from $\epsilon_\mathrm{rms} = \epsilon_\mathrm{n, rms} / \overline{p}_z $ where $\overline{p}_z$ is the average longitudinal momentum normalised to $m_e c$.

Figure \ref{Fig:Emittance}(a) shows that $\epsilon_\mathrm{rms}$ increases  rapidly  until $z = \unit[200]{\mu m}$ whilst new nitrogen electrons are ionized and  some of these slip backwards in the wakefield. After the intensity of the injection pulse has reduced sufficiently for ionization of  N$^{\geq 5+}$  to stop, $\epsilon_\mathrm{rms}$ decreases approximately as $1 / \gamma$ --- where $\gamma$ is the mean relativistic factor of the trapped electrons ---  which is consistent with the normalised emittance of the bunch not varying significantly as it is accelerated. The final normalized emittance reaches $\unit[2.0]{\mu m}$ at $z \approx \unit[6.0]{mm}$, close to the point of maximum electron energy. The rms energy spread  $\Delta E_\mathrm{rms}$ of electrons ionized from N$^{\geq 5+}$ initially increases  rapidly as the number of these electrons increases, before increasing more slowly and then decreasing as the electron bunch rotates in phase space. The relative energy spread decreases even before ionization ceases, since it is dominated by the increase in $\gamma$; a minimum value of $\Delta E_\mathrm{rms} / E = 2\%$ is reached just after the point of dephasing.

The effect of using longer drive and injection pulses is shown in Figure \ref {Fig:Emittance}(b, d). The main difference is that the interaction of the bunch with the tail of the injection pulse causes  the transverse emittance  to increase before the dephasing point is reached. As a result a minimum value of $\epsilon_\mathrm{n, rms} = \unit[2.0]{\mu m}$ is reached at $z = \unit[1.0]{mm}$; this increases to $\epsilon_\mathrm{n, rms} = \unit[5]{\mu m}$ at the point of minimum relative energy spread.
 
In summary, we have proposed a new scheme for controlling the injection and trapping of electrons into a quasi-linear  laser-driven wakefield. Non-optimized numerical simulations  show that this method can generate electron bunches of charge $\unit[5]{p C}$, mean energy \unit[370]{MeV}, relative energy spread of 2\%, and a normalized emittance of $\unit[2.0]{\mu m}$.

We note that the energy spread might be reduced still further by employing simultaneous space-time focussing (SSTF) \cite{Zhu:2005, Durfee:2012} for the injection pulse, which could substantially reduce the  distance over which the intensity of the pulse remains high. In practice reduction of the depth of focus by an order of magnitude is possible, and additional control can be achieved by combining SSTF with Bessel beam generation \cite{Clerici:2010}. These methods would further constrain the region in which injection occurs, without reducing the spot size of the injection pulse to unpractically small values or to values for which the trapped charge is small.

\begin{acknowledgments}
This work was supported by the Engineering and Physical Sciences Research Council (grant no.\ EP/H011145/1) and by The Leverhulme Trust (grant no.\ F/08 776/G). The authors would like to acknowledge the OSIRIS Consortium, consisting of UCLA, IST (Lisbon, Portugal), and USC, for the use of OSIRIS, and IST for providing access to the OSIRIS 2.0 framework. We also acknowledge helpful discussions with R.M.G.M. Trines and are grateful for computing resources provided by STFC's e-Science facility.
\end{acknowledgments}

\bibliography{References}

\end{document}